\documentclass[aps,prl,twocolumn,nofootinbib,
superscriptaddress,
]{revtex4-1}

\usepackage{hyperref}
\usepackage{url}
\usepackage{setspace}
\usepackage{amsmath,amssymb,amscd,braket}
\usepackage{amsfonts}
\usepackage{color}
\usepackage{graphicx}
\usepackage{xcolor}

\newcommand{\re}{\mathfrak{Re}}

\newcommand{\bea}{\begin{eqnarray}}
\newcommand{\eea}{\end{eqnarray}}
\newcommand{\be}{\begin{equation}}
\newcommand{\ee}{\end{equation}}
\newcommand{\ba}{\begin{align}}
\newcommand{\ea}{\end{align}}

\newcommand{\FIV}[9]{\mathfrak{F}\bigg( \begin{matrix} #2 \\ #1 \end{matrix} \, #3 \, \begin{matrix} #4  \\ \, \end{matrix} \,#5 \, \begin{matrix} #6 \\ #7 \end{matrix} ; #8, #9 \bigg)}

\newcommand{\FIVsc}[9]{\mathcal{F}\bigg( \begin{matrix} #2 \\ #1 \end{matrix} \, #3 \, \begin{matrix} #4  \\ \, \end{matrix} \,#5 \, \begin{matrix} #6 \\ #7 \end{matrix} ; #8, #9 \bigg)}

\begin{document}
\title{
Exact low-temperature Green’s functions in AdS/CFT:\\
From Heun to confluent Heun
}

\author{Paolo Arnaudo}
\email{p.arnaudo@soton.ac.uk}
\affiliation{Mathematical Sciences and STAG Research Centre, University of Southampton, Highfield, Southampton SO17 1BJ, UK}

\author{Benjamin Withers}
\email{b.s.withers@soton.ac.uk}
\affiliation{Mathematical Sciences and STAG Research Centre, University of Southampton, Highfield, Southampton SO17 1BJ, UK}

\begin{abstract}
We obtain exact expressions for correlation functions of charged scalar operators at finite density and low temperature in CFT$_4$ dual to the RN-AdS$_5$ black brane. We use recent developments in the Heun connection problem in black hole perturbation theory arising from Liouville CFT and the AGT correspondence. The connection problem is solved perturbatively in an instanton counting parameter, which is controlled in a double-scaling limit where $\omega, T \to 0$ holding $\omega/T$ fixed. This provides analytic control over the emergence of the zero temperature branch cut as a confluent limit of the Heun equation. From the Green's function we extract analytic results for the critical temperature of the holographic superconductor, as well as dispersion relations for both gapped and gapless low temperature quasinormal modes. We demonstrate precise agreement with numerics.
\end{abstract}

\maketitle

\section{Introduction and results}
The real-time linear response of strongly coupled quantum field theories (QFTs) at finite temperature and density is a challenging domain where the techniques of holographic duality \cite{Maldacena:1997re, Witten:1998qj} can be exploited to perform first-principles computations. In this context, retarded correlators are of considerable interest in the context of condensed matter physics \cite{Hartnoll:2009sz, AdSCMTbook}, nuclear collisions \cite{Policastro:2001yc, Romatschke:2017ejr, Berges:2020fwq}, and neutron stars \cite{Kovensky:2021kzl, Hoyos:2021uff}. 

In this work we analytically compute retarded correlation functions at finite density and low temperature by exploiting recent developments connecting black hole perturbation theory to Seiberg-Witten theory \cite{Aminov:2020yma}, and subsequently the connection problem for the Heun equation \cite{Bonelli:2021uvf, Bianchi:2021xpr, Bonelli:2022ten, Lisovyy2022}, which governs the linear response problem in holography. We gain unprecedented insight into the analytic structure of low-temperature correlators, including an exact description of the coalescence of poles into branch cuts at zero temperature, as well as analytic control over pole locations and the superconducting phase transition. Other recent works in the holographic context that have utilised such techniques are \cite{Dodelson:2022yvn, Jia:2024zes}. 

The focus of this work are retarded Green's functions of scalar operators of conformal dimension $\Delta$,
\bea
G^R(x,y) = -i\theta(x^0-y^0)\left<[{\cal O}_\Phi(x),{\cal O}_\Phi(y)]\right>,\label{GreensDef}
\eea
in momentum space, $\tilde{G}^R(p) = \int d^4x G^R(0,x) e^{-i p x}$ with $p^\mu = (\omega,\vec{k})$. We focus on the $3+1$-dimensional QFT dual to the Reissner-Nordstr\"om (RN)-AdS$_5$ black brane at chemical potential $\mu$, as a solution of the classical equations of motion of the action, $S = \int \sqrt{-g}d^5x(R- F^2 + 12)$, where $F=dA$, given by the following spacetime metric,
\bea
ds^2 &=& -f(r) dt^2 + \frac{dr^2}{f(r)} + r^2 d\vec{x}^2,\\
f(r) &=& r^2 - \frac{1 + \frac{4\mu^2}{3}}{r^2} + \frac{4\mu^2}{3r^4}
\eea
and gauge field,
\be
A = \mu \left(1 - \frac{1}{r^2}\right)dt.
\ee
Here the AdS radius is one, the event horizon is located at $r=1$ and the temperature $T = \frac{3-2\mu^2}{3\pi}$.

\begin{figure}[]
    \centering
    \includegraphics[width=\columnwidth]{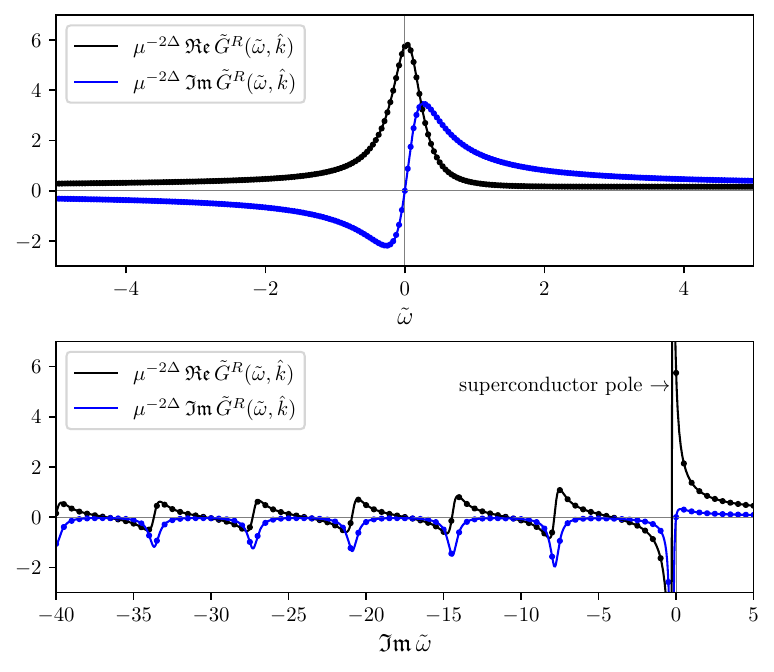}
    \caption{Comparison of analytic and numerical results for the retarded Green's function. Solid curves show the analytic expressions \eqref{Greensexpr} while dots are numerics. We take $\Delta = 5/2$, $\hat{k} = 0$ and $\nu^2 = -0.07$ at $\hat{T} = 4\times 10^{-4}$, where $\hat{T}_c = 3.4\times 10^{-4}$. The visible peak is the imprint of the superconductor zero mode since we are near $\hat{T}_c$. Here $N_\rho = 400$.  \textbf{Upper panel:} Along the real $\tilde{\omega}$ axis. \textbf{Lower panel:} Along the imaginary $\tilde{\omega}$ axis.}
    \label{fig:numerics_greens}
\end{figure}

To compute \eqref{GreensDef} in this state we consider linearised charged scalar field perturbations $\Phi$ on this background, dual to the scalar operator ${\cal O}_\Phi$. $\Phi$ obeys $D^2\Phi = m_\Phi^2\Phi$ where $D \equiv \nabla - i e A$ and $m_\Phi^2 = \Delta(\Delta-4)$. We restrict to $\Delta > 2$ and present our method at generic values of $\Delta$.

Our results are as follows. We denote variables made dimensionless with $\mu$ using hats, i.e. $\hat{T} = \frac{T}{\mu}$, $\hat{k} = \frac{k}{\mu}$, $\hat{\omega} = \frac{\omega}{\mu}$ and we consider the double scaling regime $\hat{T}\to 0$ and $\hat{\omega}\to 0$ holding $\tilde{\omega} = \frac{\omega}{T}$ fixed. We compute $\tilde{G}^R(\tilde{\omega}, \hat{k})$, which takes the following form
\be\label{Greensexpr}
\tilde{G}^R(\tilde{\omega}, \hat{k}) = N\frac{\mathcal{C}_{4-\Delta,\nu}(\tilde{\omega},\hat{k}) - \mathcal{C}_{4-\Delta,-\nu}(\tilde{\omega},\hat{k})\, {\cal G}(\tilde{\omega}, \hat{k})}{\mathcal{C}_{\Delta,\nu}(\tilde{\omega},\hat{k})-\mathcal{C}_{\Delta,-\nu}(\tilde{\omega},\hat{k})\,{\cal G}(\tilde{\omega}, \hat{k})},
\ee
where the various terms appearing $N, \mathcal{C}_{\Delta,\nu}, \mathcal{G}$ are determined analytically as a systematic expansion in $\hat{T}$ within the double scaling regime,
\be
\mathcal{C}_{\Delta,\nu}(\tilde{\omega},\hat{k}) = \mathcal{C}^{(0)}_{\Delta,\nu}(\hat{k})+\mathcal{C}^{(1)}_{\Delta,\nu}(\tilde{\omega},\hat{k})\,\hat{T}+\mathcal{O}\left(\hat{T}^2\right),
\ee
\bea
{\cal G}(\tilde{\omega},\hat{k})  &=& {\cal G}^{(0)}(\tilde{\omega},\hat{k})\hat{T}^{2\nu}+{\cal G}^{(1,+)}(\tilde{\omega},\hat{k}) \hat{T}^{2\nu+1}\log\hat{T}\nonumber\\
&&+ {\cal G}^{(1,-)}(\tilde{\omega},\hat{k}) \hat{T}^{2\nu+1} + \mathcal{O}\left(\hat{T}^{2\nu + 2}\log{\hat{T}}\right)
\eea
\be
N = 2(\Delta-2)\, 3^{\Delta -2} + \mathcal{O}(\hat{T}).
\ee
We present explicit expressions for the contributions appearing in \eqref{Greensexpr}: $\mathcal{C}^{(0)}_{\Delta,\nu}(\hat{k})$ \eqref{calC0}, $\mathcal{C}^{(1)}_{\Delta,\nu}(\tilde{\omega},\hat{k})$ \eqref{C1}, ${\cal G}^{(0)}(\tilde{\omega},\hat{k})$ \eqref{calG0}, ${\cal G}^{(1,\pm)}(\tilde{\omega},\hat{k})$ \eqref{G1+}-\eqref{G1-}. One may straightforwardly extend this series to higher orders using the methodology presented.

In particular note that ${\cal G}^{(0)}\hat{T}^{2\nu}$ can be interpreted as a finite-$\hat{T}$ AdS$_2$ Green's function, where
\bea
&&{\cal G}^{(0)}(\tilde{\omega}, \hat{k}) =\left(\frac{2\pi^2}{27}\right)^\nu \frac{\Gamma(-2\nu)^2}{\Gamma(2\nu)^2} \label{calG0}\\
&& \qquad\qquad\times  \frac{\Gamma\left(\frac{1}{2} - \frac{i e}{2\sqrt{6}} + \nu \right)}{\Gamma\left(\frac{1}{2} - \frac{i e}{2\sqrt{6}} - \nu \right)}  \frac{\Gamma\left(\frac{1}{2}+\nu + \frac{ie}{2\sqrt{6}} - \frac{i \tilde{\omega}}{2\pi}\right)}{\Gamma\left(\frac{1}{2}-\nu + \frac{ie}{2\sqrt{6}} - \frac{i \tilde{\omega}}{2\pi}\right)}, \nonumber
\eea
and where the parameter
\be
\nu \equiv \frac{1}{2}\,\sqrt{1 + \frac{\Delta(\Delta-4)}{3} + \frac{\hat{k}^2}{2} - \frac{e^2}{6}}\label{nuplanar}
\ee
labels representations of the near-horizon AdS$_2$ theory at $\hat{T}=0$ with conformal dimensions $\delta_{\pm} = \frac{1}{2}\pm \nu$  \cite{Faulkner:2009wj}. In particular, $\tilde{G}^R$ is invariant under $\nu \to -\nu$. Some parts of this analytic structure were seen in low $\hat{\omega}$ perturbation theory near extremality \cite{Faulkner:2009wj, Faulkner:2011tm, Hartnoll:2012rj, Ren:2012hg}. In \cite{Faulkner:2009wj, Faulkner:2011tm, Hartnoll:2012rj} the structure seen in \eqref{calG0} was obtained as a correlator for a black hole inside AdS$_2$. In \cite{Ren:2012hg} the leading matching coefficients for a small-$\hat{\omega}$ expansion were obtained using hypergeometric connection formulae which arise at extremality, and we find agreement in that limit.\footnote{We note $q_\text{there} = e_\text{here}/2$, $T_\text{there} =\sqrt{3/2} \hat{T}_\text{here}$ and ${\cal G}_{k,\text{there}}^{(T)} = 2^{2\nu}3^{4\nu} \Gamma(2\nu) {\cal G}^{(0)}_\text{here}/\Gamma(-2\nu)$. } See also \cite{Davison:2013bxa, Arean:2020eus, Ren:2020djc, Wu:2021mkk, Ren:2021rhx, Davison:2022vqh} for other related low $\hat{\omega}, \hat{T}$ matching computations involving AdS$_2$ critical points. Here we emphasise that the Heun connection problem is naturally controlled by low $\hat{T}$ at fixed $\tilde{\omega} = \omega/T = \mathcal{O}(1)$, which provides a systematic treatment of the correlators -- indeed, $\tilde{\omega}$ appears non-trivially inside each of ${\cal G}^{(0)}(\tilde{\omega},\hat{k})$, $\mathcal{C}^{(1)}_{\Delta,\nu}(\tilde{\omega},\hat{k})$, ${\cal G}^{(1,\pm)}(\tilde{\omega},\hat{k})$.

\begin{figure}[]
    \centering
    \includegraphics[width=\columnwidth]{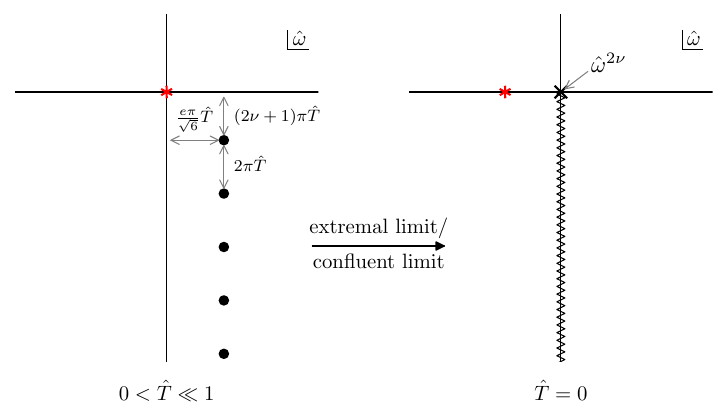}
    \caption{The analytic structure of $\tilde{G}^R$ at low ${\hat T} \sim \hat{\omega}$ in the $\hat{\omega}$ plane, described using closed-form solutions to the Heun connection problem. The analytic expression for $\tilde{G}^R$ \eqref{Greensexpr} captures the process where the line of poles \eqref{qnms} coalesces in to a branch cut at $\hat{T} = 0$, as a confluent limit of the Heun equation via \eqref{stirling}. Optionally, by tuning model parameters so that the holographic superconductor critical temperature $\hat{T}_c \ll 1$, then \eqref{Greensexpr} also captures the holographic superconductor zero mode \eqref{hydromode},\eqref{superconductormode} (red asterisk) and $\hat{T}_c$ itself.}
    \label{fig:sketch}
\end{figure}

The closed-form Green's function \eqref{Greensexpr} is the full RN-AdS$_5$ correlator -- rather than an AdS$_2$ near-horizon result -- and it displays an intricate analytic structure for which we find excellent agreement with low-temperature numerics, as shown in figure \ref{fig:numerics_greens}. Moreover, we note that taking ${\hat T} \to 0$ holding $\hat{\omega}$ fixed one has
\begin{equation}\label{stirling}
\hat{T}^{2\nu}\frac{\Gamma\left(\frac{1}{2}+\nu + \frac{ie}{2\sqrt{6}} - \frac{i \hat{\omega}}{2\pi \hat{T}}\right)}{\Gamma\left(\frac{1}{2}-\nu + \frac{ie}{2\sqrt{6}} - \frac{i \hat{\omega}}{2\pi \hat{T}}\right)}\to \left(- \frac{i \hat{\omega}}{2\pi }\right)^{2\nu},
\end{equation}
where we used the Stirling approximation $\Gamma(z+\alpha)/\Gamma(z)\sim z^{\alpha}$ for $z\to\infty$ and $\alpha\in\mathbb{C}$, and then took the $\hat{T}\to 0$ limit. This analytically describes a line of poles coalescing into a branch point at $\hat{\omega} = 0$, capturing the process previously observed numerically in \cite{Edalati:2009bi, Edalati:2010hk, Edalati:2010pn} in the context of RN-AdS$_4$. This is illustrated in figure \ref{fig:sketch}.

When $\re(\nu) > 0$ the line of poles in question can be determined analytically. Because of the presence of the term $\hat{T}^{2\nu}$, these poles can be read off from the poles of the $\Gamma$-function in the numerator of $\mathcal{G}^{(0)}$,
\bea
\tilde{\omega} &=& \tilde{\omega}_0 + \tilde{\omega}_1 \hat{T} + \tilde{\omega}_{2\nu} \hat{T}^{2\nu} + O(\hat{T}^2, \hat{T}^{4\nu}),\label{qnms}\\
\tilde{\omega}_0 &=& \frac{\pi e}{\sqrt{6}} - i \pi(2\nu + 2n + 1)\label{omega0}
\eea
where $n\in \mathbb{Z}_{\geq 0}$ and where $\tilde{\omega}_{2\nu}$ and $\tilde{\omega}_1$ are given in \eqref{omega2nu} and \eqref{omega1} respectively.\footnote{When $e=0$ the leading order term $\tilde{\omega}_0$ was previously obtained in \cite{Jia:2024zes} using similar methods.}. A comparison of these terms to numerics is presented in table \ref{tab:numerical_qnms_fundamental}.

\begin{figure}[]
    \centering
    \includegraphics[width=\columnwidth]{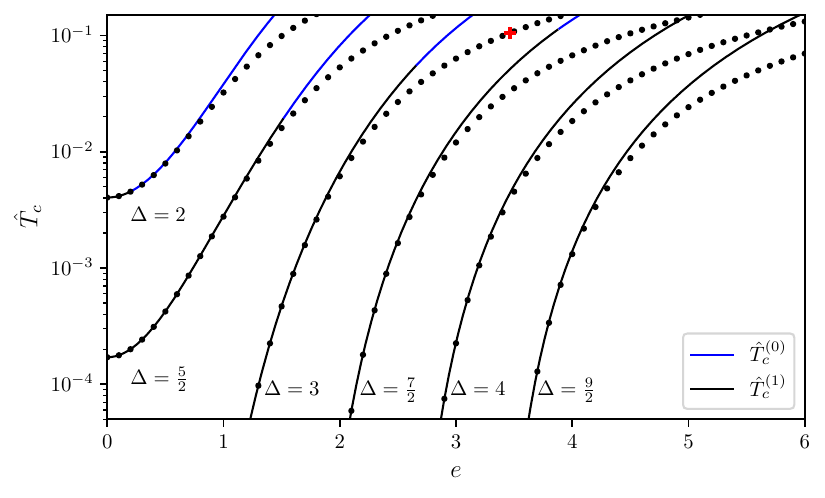}
    \caption{Comparison of analytic and numerical results for the holographic superconductor critical temperature at various $\Delta$.
    Solid curves show the analytic result \eqref{crittemp_n}, while dots are numerics. For \eqref{crittemp_n} to apply, $\Delta, e$ must be such that $\hat{T}_c$ is small, and in this regime we find excellent agreement. The red plus corresponds to the top-down superconductor of \cite{Gubser:2009qm} which we discuss further in the conclusions.
    }
    \label{fig:numerics_tc}
\end{figure}

Additionally, using \eqref{Greensexpr} we gain analytic control over the phase transition of the holographic superconductor \cite{Gubser:2008px, Hartnoll:2008vx, Hartnoll:2008kx}. If parameters are chosen such that $\hat{T}_c$ for superconductivity is low, then we can obtain $\hat{T}_c$ and the associated zero mode dispersion relation from \eqref{Greensexpr}. For generic $\nu$, $\hat{T}_c$ assumes the $n$-sheeted expression\footnote{Strictly speaking $\hat{T}_c$ here corresponds to the temperature at which a zero mode appears in the spectrum.}
\begin{equation}\label{crittemp_n}
\hat{T}_c^{(n)}=\exp\left[\frac{1}{2\tilde{\nu}}\log\left(\frac{\mathcal{C}^{(0)}_{\Delta,\tilde{\nu}}(0)}{\mathcal{C}^{(0)}_{\Delta,-\tilde{\nu}}(0)\mathcal{G}^{(0)}(0,0)}\right)-\frac{i\,\pi\, n}{\tilde{\nu}}\right],
\end{equation}
where $\tilde{\nu} = \nu|_{\hat{k}=0}$.
When $\re(\nu)=0$, a superconducting instability is expected on the grounds that the AdS$_2$ BF-bound is violated \cite{Gubser:2008px}. For $0 < 2\nu \leq 1$, there is a single real sheet, given by $\hat{T}_c^{(0)}$, but for $\re(\nu)=0$ there are multiple sheets labeled by $n$. As we discuss shortly, the relevant sheets at asymptotically low temperature correspond to $n=1$, and extending to higher temperature by continuity. The other physical sheets correspond to subleading instabilities as higher overtone QNMs cross into the upper half plane. See figure \ref{fig:numerics_tc} where we compare $\hat{T}_c^{(n)}$ to numerics. Finally, at $\hat{T} = \hat{T}_c^{(n)}$, we extract the hydrodynamic dispersion relation for the superconductor zero mode, 
\be\label{hydromode}
\hat{\omega}^{(n)}(\hat{k}) = -i\, D^{(n)}\, \hat{k}^2 + O(\hat{k}^4)
\ee
with $D^{(n)}$ given in \eqref{hydroorder}. We remark that while the gapped modes \eqref{omega0} at leading order are associated with a near-horizon AdS$_2$ at $\hat{T} = 0$, the mode \eqref{hydromode} is not.

\section{Method and Analysis}
The first step is to relate the equation of motion for $\Phi$ to the Heun equation. We begin with a new radial variable 
\be \label{zdef}
z = \frac{\sqrt{48 \mu ^2+9}+9}{6 \left(1-r^2\right)},
\ee
with $z = 0$ corresponding to the conformal boundary and $z=\infty$ to the black hole horizon. Under separation of variables and an appropriate choice of prefactor\footnote{The redefinition of the wave function using this prefactor can be reinterpreted as a particular choice of spacetime foliation as discussed in \cite{Minucci:2024qrn}.},
\begin{equation}
\begin{aligned}
\Phi &=e^{-i \omega t + i \vec{k}\cdot \vec{x}} \frac{\sqrt{z} }{\sqrt{1-z} \sqrt[4]{6 z^2-4 \mu ^2 (z+1)^2-15 z+6}}\\
&\times\exp \left(-\frac{1}{2} \tanh ^{-1}\left(\frac{8 \mu ^2+4 \left(2 \mu ^2-3\right) z+15}{3 \sqrt{48 \mu ^2+9}}\right)\right)\psi(z)
\end{aligned}
\end{equation}
then $(D^2-m_\Phi^2)\Phi = 0$ reduces to the Heun equation
\begin{widetext}
\begin{equation}\label{heunnormalform}
\psi''(z) + \left[\frac{\frac{1}{4}-a_0^2}{z^2}+\frac{\frac{1}{4}-a_1^2}{(z-1)^2} + \frac{\frac{1}{4}-a_t^2}{(z-t)^2}- \frac{\frac{1}{2}-a_1^2 -a_t^2 -a_0^2 +a_\infty^2 + u}{z(z-1)}+\frac{u}{z(z-t)} \right]\psi(z)=0,
\end{equation}
\end{widetext}
with regular singular points at $0, 1, t, \infty$. The parameters $a_0, a_1, a_t, a_\infty, u$ and singular point $t$ are given in the supplemental material \eqref{dictiogauge}.
We remark that $1/t$ is proportional to $\hat{T}$:
\begin{equation}\label{tandTexpansions}
\frac{1}{t}=\frac{2\pi}{9}\hat{T}+\mathcal{O}\left(\hat{T}^2\right),
\end{equation}
so that in the extremal limit the singular point $t$ collides with the one at infinity, resulting in the confluent Heun equation with an irregular singular point. The local solutions around the singularities of the differential equation \eqref{heunnormalform} can be written in terms of the Heun function \cite{heun1888theorie,ronveaux1995heun}
\begin{equation}
\mathrm{Heun}(t,q;\alpha,\beta,\gamma,\delta;z)=1+\frac{q}{t\,\gamma}\,z+\mathcal{O}\left(z^2\right),
\end{equation}
by considering the wave function
\begin{equation}
w(z)=z^{-\frac{1}{2}+a_0} (1-z)^{-\frac{1}{2}+a_1} (t-z)^{-\frac{1}{2}+a_t}\,\psi(z).
\end{equation}
In particular, we are interested in the basis of solutions around $z=\infty$, the horizon, and around $z=0$, the AdS boundary. At $z=\infty$ we require ingoing perturbations for retarded Green's functions \cite{Son:2002sd, Herzog:2002pc, vanRees:2009rw}, which selects the behaviour,
\begin{equation}
w^{(\infty)}(z)\sim z^{-1+a_0+a_1+a_t+a_{\infty}}\quad\text{for}\ z\sim\infty.
\end{equation}
This can be analytically continued to the region close to $z=0$ where it can be written as
\begin{equation}\label{connectionformula}
\mathfrak{C}_+^{(\infty\,0)}w_+^{(0)}(z)+\mathfrak{C}_-^{(\infty\,0)}w_-^{(0)}(z),
\end{equation}
where the independent solutions at the boundary $w_{\pm}^{(0)}(z)$ admit the behavior
\bea
w_-^{(0)}&\sim& 1\quad \text{for}\ z\sim 0,\\
w_+^{(0)}&\sim& z^{2a_0}\quad \text{for}\ z\sim 0.
\eea
From \eqref{connectionformula} the retarded Green's function can be read off,
\be
\tilde{G}^R(\tilde{\omega}, \hat{k}) = \mathcal{N} \frac{\mathfrak{C}_+^{(\infty\,0)}}{\mathfrak{C}_-^{(\infty\,0)}},
\ee
where $\mathcal{N}$ accounts for additional constant factors picked up in a change of variables from $z$ as defined in \eqref{zdef} to Fefferman-Graham coordinates, and constant prefactors from holographic renormalisation \cite{Skenderis:2002wp},
\bea
&& \mathcal{N} = 2(\Delta-2) \label{Ndef}\\
&&\times \left(-\frac{3}{2} - \frac{1}{2}\sqrt{9 + 2\pi \hat{T}\left(3\pi \hat{T} - \sqrt{24 + 9\pi^2\hat{T}^2}\right)}\right)^{\Delta -2}.\nonumber
\eea

The connection formulas for $\mathfrak{C}_{\pm}^{(\infty\,0)}$ in \eqref{connectionformula} were obtained in \cite{Bonelli:2022ten} by studying the connection coefficients of five-point conformal blocks with a degenerate insertion in Liouville conformal field theory.\footnote{An alternative approach can be taken using recurrence relations, as discussed in \cite{Lisovyy2022}. In this work we follow the approach of \cite{Bonelli:2021uvf,Bonelli:2022ten}.} These satisfy the BPZ equation \cite{Belavin:1984vu}, which in the semiclassical limit becomes a Heun equation. The outcome is, 
\bea\label{connectioncoeff}
\mathfrak{C}_{\pm}^{(\infty\,0)}=\sum_{\sigma=\pm}\biggl[ \frac{\Gamma(1-2\sigma a)\Gamma(-2\sigma a)\Gamma(1-2a_\infty)\Gamma(\mp 2a_{0})}{\prod_{\theta=\pm}\Gamma\left(\frac{1}{2}-a_{\infty}+\theta a_t-\sigma a\right)}\nonumber\\
\times\frac{e^{-\frac{1}{2}\left(\pm \partial_{a_0} + \partial_{a_\infty} + \sigma \partial_a\right) F(1/t)+i \pi\left( 1-a_1\mp a_0\right)}}{t^{\frac{1}{2}-a_{\infty}-a_t+\sigma a}\prod_{\theta=\pm}\Gamma\left(\frac{1}{2}\mp a_{0}+\theta a_1-\sigma a\right)}\biggr].
\eea
The analysis to get this result proceeds in two steps: considering the pants-decomposition of the four-punctured sphere representing the $z$-space of the Heun equation, $z=0$ and $z=\infty$ lie in two different patches. To connect them, one has to pass through the intermediate region, whose monodromy properties are encoded in the parameter $a$. $a$ is related to the instanton part of the Nekrasov-Shatashvili free-energy, $F(1/t)$, through  Matone relation \cite{Matone:1995rx},
\begin{equation}
u =-\frac{1}{4} + a^2 - a_{\infty}^2 + a_t^2 + t\, \partial_t F(1/t). \label{Matone}
\end{equation}
Further details of this procedure and of the computation of $F(1/t)$ are given in the supplemental material. At leading order one has
\be\label{NSfirstinstanton}
F(1/t)=\frac{\left( a^2-a_0^2+ a_1^2-\frac{1}{4}\right) \left(a^2- a_{\infty}^2+a_{t}^2-\frac{1}{4}\right)}{\left(\frac{1}{2}-2 a^2\right)\,t}+\mathcal{O}\left(\frac{1}{t^2}\right)
\ee
Then $a$ can be obtained by inverting \eqref{Matone} order-by-order in $1/t$. At leading order in the temperature, provided \eqref{tandTexpansions}, one has $a = \nu + O(\hat{T})$ with $\nu$ given in \eqref{nuplanar}.
The final step to obtain \eqref{Greensexpr} is to note that the subleading corrections in $\hat{T}$ come with prefactors that behave as $\hat{\omega}/\hat{T}$. If $\hat{\omega}$ is treated as $O(\hat{T}^0)$ then these would appear at leading order in the $\hat{T}$ expansion, requiring resummation. Instead, here we hold $\tilde{\omega} \equiv \hat{\omega}/\hat{T}$ fixed which leads to a controlled $\hat{T}$ expansion. This completes the derivation of \eqref{Greensexpr}.

Finally we present some additional details in the analysis of the resulting Green's function \eqref{Greensexpr}. The pole locations \eqref{qnms} can be obtained analytically by distinguishing different regimes for the parameter $\nu$.
Let us first consider the case $\re(\nu)>0$. Because of the presence of the term $\hat{T}^{2\nu}$, the leading order of $\tilde{\omega}$ in $\hat{T}$ is found by looking at the poles of the $\Gamma$-function in the numerator of $\mathcal{G}^{(0)}$ including $\tilde{\omega}$ in its argument:
\begin{equation}
\frac{1}{2}+\nu+\frac{i e}{2 \sqrt{6}} -\frac{i \tilde{\omega}_0}{2 \pi }=-n,\quad n\in\mathbb{Z}_{\ge 0}
\end{equation}
from which we obtain \eqref{omega0}. The first subleading correction depends on the value of $\re(2\nu)$. If $0<\re(2\nu)\le 1$, then setting
$\tilde{\omega}=\tilde{\omega}_0+\tilde{\omega}_{2\nu}\hat{T}^{2\nu}$ and plugging this in $\mathcal{G}^{(0)}$ gives
\begin{equation}
\begin{aligned}
\mathcal{G}^{(0)}\hat{T}^{2\nu}=&\left(\frac{2\pi^2}{27}\right)^{\nu}\frac{2 i \pi  (-1)^n}{\tilde{\omega}_{2\nu}\,\Gamma(-n-2\nu)\,  n!}\\
&\times\frac{\Gamma (-2 \nu )^2\Gamma \left(\frac{1}{2}+\nu-\frac{i e}{2 \sqrt{6}} \right)}{\Gamma (2 \nu )^2 \Gamma \left(\frac{1}{2}-\nu-\frac{i e}{2 \sqrt{6}} \right)}+\mathcal{O}\left(\hat{T}^{2\nu}\right),
\end{aligned}
\end{equation}
from which
\bea\label{omega2nu}
\tilde{\omega}_{2\nu}&=&\left(\frac{2\pi^2}{27}\right)^{\nu}\frac{2 i \pi  (-1)^n}{\Gamma(-n-2\nu)\,  n!}\frac{\Gamma \left(\frac{1}{2}+\nu-\frac{i e}{2 \sqrt{6}} \right)}{ \Gamma \left(\frac{1}{2}-\nu-\frac{i e}{2 \sqrt{6}} \right)}\nonumber\\
&\times&\frac{\Gamma (-2 \nu )^2\prod_{\sigma=\pm}\Gamma \left(\nu+\sigma\,\frac{e}{2 \sqrt{3}} +\frac{\Delta-1}{2}\right)}{\Gamma (2 \nu )^2\prod_{\sigma=\pm}\Gamma \left(-\nu+\sigma\,\frac{e}{2 \sqrt{3}} +\frac{\Delta-1}{2}\right)}.
\eea
The coefficients of the corrections $\tilde{\omega}_k\hat{T}^k$, $k=1,\dots,\lfloor\re(2\nu)\rfloor$ in the $\tilde{\omega}$-expansion can be obtained by looking at the poles of the $\Gamma$-function $\Gamma\left(\frac{1}{2}-a_{\infty}+a_t+a\right)$ by considering the $1/t$ expansion of $a$ up to order $\lfloor\re(2\nu)\rfloor$. This is the generalization of the procedure providing the expression of $\tilde{\omega}_0$ seen above. For example, the expression of $\tilde{\omega}_1$ reads
\begin{equation}\label{omega1}
\begin{aligned}
&\tilde{\omega}_1=\frac{\pi ^2}{36}\Biggl\{\frac{(2 \nu+2 n+1) \left[3 e \left(4 \Delta(\Delta-4)+9 \hat{k}^2-8\right)-7 e^3\right]}{ 12 \nu \left(4 \nu^2-1\right)}\\
&+i \left[\frac{7 e^2-14(\Delta-1)(\Delta-3)-3 \hat{k}^2}{2 \sqrt{6} \nu}-7 \sqrt{6} (2 n+1)\right]\Biggr\}.
\end{aligned}
\end{equation}
After including these integer corrections, the term $\tilde{\omega}_{2\nu}\hat{T}^{2\nu}$ in the $\tilde{\omega}$-expansion will not be affected.

Let us turn to the critical temperature result \eqref{crittemp_n}. This is obtained by setting $\tilde{\omega} = \hat{k} = 0$ and finding $\hat{T}$ at which there is a zero of the denominator of \eqref{Greensexpr} at leading order,
\begin{equation}\label{zerothGreen}
\frac{\mathcal{C}_{4-\Delta,\tilde{\nu}}^{(0)}(0) - \mathcal{C}_{4-\Delta,-\tilde{\nu}}^{(0)}(0)\, {\cal G}^{(0)}(0, 0)\hat{T}_c^{2\tilde{\nu}}}{\mathcal{C}_{\Delta,\tilde{\nu}}^{(0)}(0)-\mathcal{C}_{\Delta,-\tilde{\nu}}^{(0)}(0)\,{\cal G}^{(0)}(0, 0)\hat{T}_c^{2\tilde{\nu}}}.
\end{equation}
Since $\hat{T}$ appears only in the combination $\hat{T}^{2\tilde{\nu}}$ there are multiple solutions given by \eqref{crittemp_n}. Due to the low $\hat{T}$ expansion employed in \eqref{zerothGreen} we can only capture those branches of solution which admit a parameter regime with asymptotically low $\hat{T}_c$. Near $\nu^2 = 0^-$ the $n \geq 1$ branches have a low $\hat{T}$ regime. Near $2\nu = 1$ the $n=0$ branch also has a low $\hat{T}$ regime, however, we find no zero mode associated to it; if we expand \eqref{zerothGreen} for $\hat{T}_c = \hat{T}_c^{(0)}$ around $\nu=1/2$, we find an expansion without poles at $\nu=1/2$. Indeed, there are a pair of poles, one of either side of $\nu=1/2$, whose combined contribution to the full Green's function appears at order $\hat{T}$. Thus we expect the physical $\hat{T}_c$ is given by $n=1$ near $\tilde{\nu}^2 = 0^-$, and extends to higher $\hat{T}_c$ by continuity, along which $n$ jumps.

When considering the extremal limit $\hat{T}=0$, the generation of the branch cut in $\omega$ can be seen from \eqref{stirling}. Substituting \eqref{stirling} in the leading order of $\mathcal{G}(\hat{\omega},\hat{k})$ before rescaling the frequency with the temperature, we find the analytic result for the holographic superconductor zero mode, appearing as a red asterisk in figure \ref{fig:sketch}:
\begin{equation}\label{superconductormode}
\hat{\omega}_*=2\,i\,\pi\,\exp\left[\frac{1}{2\nu}\log\left(\frac{\mathcal{C}^{(0)}_{\Delta,\nu}(\hat{k})}{\mathcal{C}^{(0)}_{\Delta,-\nu}(\hat{k})\mathcal{G}_{\text{ext}}^{(0)}(\hat{k})}\right)-\frac{i\,\pi\,n}{\nu}\right],
\end{equation}
where
\begin{equation}
\mathcal{G}_{\text{ext}}^{(0)}(\hat{k})=\left(\frac{2\pi^2}{27}\right)^{\nu}\,\frac{\Gamma(-2\nu)^2}{\Gamma(2\nu)^2}\frac{\Gamma\left(\frac{1}{2}-\frac{i\,e}{2\sqrt{6}}+\nu\right)}{\Gamma\left(\frac{1}{2}-\frac{i\,e}{2\sqrt{6}}-\nu\right)}.
\end{equation}
The integer $n$ in \eqref{superconductormode} labels the sheet around the $\hat{\omega}$ branch cut.

By considering the hydrodynamic expansion (around small $\hat{\omega}$ and small $\hat{k}$) of the leading order in $\hat{T}$ of the denominator of \eqref{Greensexpr} at $\hat{T} = \hat{T}_c^{(n)}$,
\begin{equation}
\mathcal{C}^{(0)}_{\Delta,\nu}(\hat{k})-\mathcal{C}^{(0)}_{\Delta,-\nu}(\hat{k})\mathcal{G}^{(0)}\left(\frac{\hat{\omega}}{\hat{T}_c^{(n)}},\hat{k}\right)\left[\hat{T}_c^{(n)}\right]^{2\nu},
\end{equation}
we find the hydrodynamic zero-mode dispersion relation \eqref{hydromode}.

\section{Comparison with numerics}

To compute $\tilde{G}^R(\tilde{\omega}, \hat{k})$ numerically we write $\Phi = e^{-i\omega t + i \vec{k}\cdot \vec{x}} (1-\rho^2)^{-\frac{i \tilde{\omega}}{4\pi}} \rho^{4-\Delta} h(\rho)$ where $\rho = 1/r$, so that regularity of $h$ at $\rho = 1$ corresponds to an ingoing perturbation. We then discretise the resulting $\rho$ differential operator using Chebyshev spectral methods using $N_\rho$ points, and solve for $h(\rho)$ subject to the boundary condition $h(0) = 1$. Extracting the coefficient of $\rho^{2\Delta -4}$ at the boundary gives the Green's function. An example comparison between numerics and \eqref{Greensexpr} is shown in figure \ref{fig:numerics_greens} near the superconducting critical temperature.

To compute QNMs numerically we write $\Phi = e^{-i\omega t + i \vec{k}\cdot \vec{x}} (1-\rho)^{-\frac{i \tilde{\omega}}{4\pi}} \rho^{\frac{\Delta}{2}} h(\rho)$ where $\rho = 1/r^2$, so that regularity of $h$ at $\rho = 1$ corresponds to an ingoing perturbation while regularity of $h$ at $\rho = 0$ corresponds to a normalisable mode. We then discretise the resulting $\rho$ differential operator using Chebyshev spectral methods using $N_\rho$ points, which yields a matrix eigenvalue problem after converting to first order form. Eigenvalues are obtained using Arnoldi iteration. We compute the fundamental mode and compare to \eqref{qnms} in table \ref{tab:numerical_qnms_fundamental}, with agreement continuing to improve as $\hat{T}$ is lowered. With this method, we also numerically compute $\hat{T}_c$ by iteratively root finding for the zero mode using Newton-Raphson. This is then compared to the analytic formula \eqref{crittemp_n} in figure \ref{fig:numerics_tc}.

\begin{widetext}

\begin{table}[h!]
$\begin{array}{l|l|l|l | l}
\hat{T} & \tilde{\omega} \text{ numerics} & \tilde{\omega}_0 & \tilde{\omega}_0 + \tilde{\omega}_{2\nu} \hat{T}^{2\nu} & \tilde{\omega}_0 + \tilde{\omega}_{2\nu} \hat{T}^{2\nu} + \tilde{\omega}_1 \hat{T} \\
\hline
10^{-1} & 4.273209-4.579409 i & 2.565100-\mathbf{5}.235988 i & 2.817226-\mathbf{5}.066166 i & \mathbf{3}.783624-\mathbf{5}.026993 i\\
10^{-2} & 2.726249-5.195875 i & \mathbf{2}.565100-\mathbf{5.2}35988 i & \mathbf{2}.619419- \mathbf{5.19}9401 i & \mathbf{2.7}16058- \mathbf{5.19}5483 i\\
10^{-3} & 2.586764-5.227755 i & \mathbf{2.5}65100-\mathbf{5.2}35988 i & \mathbf{2.5}76802-\mathbf{5.228}105 i & \mathbf{2.58}6466- \mathbf{5.227}714 i\\
10^{-4} & 2.568597-5.234252 i & \mathbf{2.56}5100-\mathbf{5.2}35988 i & \mathbf{2.56}7621-\mathbf{5.2342}90 i & \mathbf{2.5685}87-\mathbf{5.23425}0 i\\
10^{-5} & 2.565740-5.235618 i & \mathbf{2.56}5100-\mathbf{5.235}988 i & \mathbf{2.565}643-\mathbf{5.23562}2 i & \mathbf{2.56573}9-\mathbf{5.235618} i
\end{array}$
    \caption{Comparing analytic QNM formulae \eqref{qnms}, \eqref{omega2nu}, \eqref{omega1} at $n=0$ with a numerical determination of the fundamental mode. Chosen parameters are $\Delta = 5/2$, $e=2$ and $\hat{k} = 7/(3\sqrt{2})$ where $\nu = 1/3$. Bold indicates agreement with numerics when rounded to that many significant figures. In the `$\tilde{\omega}$ numerics' column we used $N_\rho = 200, 200, 300, 700, 2000$ (from top to bottom) to obtain convergence to the displayed number of decimal digits.}
    \label{tab:numerical_qnms_fundamental}
\end{table}
\end{widetext}

\section{Outlook}
A natural question is whether any consistent string/M-theory embeddings of the holographic superconductor \cite{Gubser:2009qm, Gauntlett:2009dn, Gauntlett:2009bh} can be analytically described. The type-IIB truncation of \cite{Gubser:2009qm} falls into the class of models we have studied here. The operator studied there has $\Delta = 3$ and $e = 2\sqrt{3}$, with numerically determined critical temperature $\hat{T}_c \simeq 0.1051$.\footnote{The chemical potential $\mu_\text{there} = \sqrt{3}\mu_\text{here}$ and the R-charge parameter $R_\text{there} = \frac{1}{\sqrt{3}}e_\text{here}$, due to the different coefficient of $F^2$ in the action.} This operator is marked in red in figure \ref{fig:numerics_tc}. There is an appreciable difference to our leading analytic result \eqref{crittemp_n}, because $\hat{T}_c$ is not particularly small. We can however go a bit further, by including additional corrections in $\hat{T}$ to the denominator of \eqref{Greensexpr} which take the following form,
\begin{equation}\label{corrected_denominator}
    \begin{aligned}
&\mathcal{C}^{(0)}_{\Delta,\nu}+\mathcal{C}^{(1)}_{\Delta,\nu}\,\hat{T} -( \mathcal{C}^{(0)}_{\Delta,-\nu}+\mathcal{C}^{(1)}_{\Delta,-\nu}\,\hat{T})[\mathcal{G}^{(0)}\hat{T}^{2\nu}\\
&\qquad\qquad+\mathcal{G}^{(1,+)}\hat{T}^{2\nu+1}\log(\hat{T})+\mathcal{G}^{(1,-)}\hat{T}^{2\nu+1}],
\end{aligned} 
\end{equation}
with $\mathcal{G}^{(1,\pm)}$ given in supplemental material.
Root finding in \eqref{corrected_denominator} for $\hat{T}$ at $\tilde{\omega} = \hat{k} = 0$ we find an improved agreement, $\hat{T}_c\simeq 0.1053\, +0.0162 i$, albeit at the cost of adding a small imaginary part (which in turn would have to be removed by even higher order terms).

RN-AdS$_4$ is desirable for comparison with condensed matter systems, however there, the analogous differential equation has 5 singular points and is not of Heun type. One may expect that the results we obtained for RN-AdS$_5$ should be similar to RN-AdS$_4$, but with the added benefit that they are analytic.

Given the analytic result for the QNMs \eqref{qnms} and the gapless mode \eqref{hydromode} it would be interesting to try to understand how they fit with recent causality bounds on QNM dispersion relations as a function of $\hat{k}$ \cite{Heller:2022ejw, Heller:2023jtd}. One difficulty is while there is a branch point at complex $\hat{k}$, this occurs at $\nu = 0$ where the approximation used to derive the QNMs \eqref{qnms} breaks down.

\paragraph{Acknowledgements.}
\begin{acknowledgments}
We thank Javier Carballo, Blaise Gout\'eraux and Jie Ren for useful discussions. 
P.\,A. is supported by the Royal Society grant URF\textbackslash R\textbackslash 231002, `Dynamics of holographic field theories'.
B.\,W. is supported by a Royal Society University Research Fellowship and in part by the Science and Technology Facilities Council (Consolidated Grant `Exploring the Limits of the Standard Model and Beyond'). 
\end{acknowledgments}

\bibliography{biblio} 

\clearpage

\section*{SUPPLEMENTAL MATERIAL}

\subsection{Gauge theory details and conventions}

The connection formula in \eqref{connectionformula} together with the expressions of the coefficients in \eqref{connectioncoeff} are written in terms of quantities coming from Liouville conformal field theory (see \cite{Teschner:2001rv,Nakayama:2004vk} for detailed reviews of the theory), or from $\mathcal{N}=2$ $SU(2)$ gauge theory with $N_f=4$ fundamental hypermultiplets via the AGT correspondence \cite{Alday:2009aq}. 
In this Appendix, we define the relevant quantities entering \eqref{connectioncoeff} and fix the conventions used. 

The connection formulas for the Heun equation were obtained in \cite{Bonelli:2022ten} by considering the connection formulas for conformal blocks in Liouville conformal field theory. Here, we describe the basics of the argument. Let us consider a correlation function with four primary insertions $V_i$, with the additional insertion of the degenerate field $\Phi_{2,1}$,
\be\label{corrdegen}
\langle V_{\alpha_\infty}(\infty)V_{\alpha_t}(t)V_{\alpha_1}(1)\Phi_{2,1}(z)V_{\alpha_0}(0)\rangle
\ee
where $\alpha_i$ denote conformal momenta of the operators and $\alpha_{2,1}=-b-1/(2b)$, where $b$ is the coupling of the theory. The conformal dimension $\Delta_{\alpha}$ of the operator with momentum $\alpha$ is $\Delta_{\alpha}=Q^2/4-\alpha^2$, where $Q=b+1/b$.  \eqref{corrdegen} satisfies a second-order partial differential equation -- the BPZ equation \cite{Belavin:1984vu}. 
In the semiclassical limit, in which $b\to 0$ and $\alpha_i\to\infty$ with $a_i\equiv b\,\alpha_i$ finite, the BPZ equation reduces to the Heun equation in $z$, with regular singular points at the primary insertions.

In the present case of interest, the insertion location $t \propto \hat{T}^{-1}$ \eqref{tandTexpansions} so that at low $\hat{T}$ we have $0 < 1 \ll t$. Moving $z$ close to the positions of different insertions then provides three different conformal block expansions of \eqref{corrdegen}, that is, $0 < z \ll 1 \ll  t$, $0 < 1 \ll z \ll t$, and $0 < 1 \ll t \ll z$. For example, expanding the correlation function for $z\sim 0$ we write
\begin{widetext}
\begin{equation}\label{corrfunctionat0}
\begin{aligned}
&\langle V_{\alpha_{\infty}}(\infty) V_{\alpha_t} (t) V_{\alpha_1} (1) \Phi_{2,1} (z) V_{\alpha_0}(0) \rangle = \sum_{\theta = \pm} \int d \alpha \, C_{\alpha_{2,1} \alpha_0}^{\alpha_{0 \theta}} C_{\alpha_1  \alpha_{0 \theta}}^{\alpha} C_{\alpha_\infty \alpha_t \alpha} \bigg|\FIV{\alpha_\infty}{\alpha_t}{\alpha}{\alpha_1}{{\alpha_{0 \theta}}}{\alpha_{2,1}}{\alpha_0}{\frac{1}{t}}{z} \bigg|^2,
\end{aligned}
\end{equation}
\end{widetext}
where $C_{\alpha_1\alpha_2\alpha_3}$ denote the 3-point functions, $C^{\alpha_1}_{\alpha_2\alpha_3}$ are OPE coefficients, and $\mathfrak{F}$ is the conformal block. We refer to \cite{Bonelli:2022ten} for detailed definitions. Crossing symmetry relates the three conformal block expansions. Used in turn, these relations connect the expansions near $z\sim 0$ to the intermediate region $1 \ll z \ll t$ and then finally to $z\sim \infty$, giving connection formulae for connecting $\mathfrak{F}$ expanded for $z \sim 0$ to $\mathfrak{F}$ expanded for $z \sim \infty$, in terms of the 3-point functions and OPE coefficients.

The corresponding Heun connection coefficients are given by taking the semiclassical limit. The conformal blocks diverge as $b\to 0$, but, the leading divergence is independent of $z$. For example, for the block in \eqref{corrfunctionat0},
\begin{equation}\label{blockat0expansion}
\begin{aligned}
&\FIV{\alpha_\infty}{\alpha_t}{\alpha}{\alpha_1}{{\alpha_{0 \theta}}}{\alpha_{2,1}}{\alpha_0}{\frac{1}{t}}{z} =\\
&t^{\Delta_{\alpha_\infty} - \Delta_{\alpha_t} - \Delta_{\alpha}} z^{\frac{b\,Q}{2} + \theta b \alpha_0} \exp \left[ \frac{1}{b^2} \left( F\left(\frac{1}{t}\right) + \mathcal{O} (b^2,z)  \right) \right] \,,
\end{aligned}
\end{equation}
where $F(1/t)$ is the classical 4-point conformal block.
Consequently, one can remove the divergence in $b\to 0$ by normalising with respect to the conformal block without the degenerate insertion, given by
\begin{equation}
\mathfrak{F} \left( \begin{matrix} \alpha_t \\ \alpha_\infty \end{matrix} \alpha \, \begin{matrix} \alpha_1 \\ \alpha_0 \end{matrix} ; \frac{1}{t} \right) = t^{\Delta_{\alpha_\infty} - \Delta_{\alpha_t} - \Delta_{\alpha}} e^{b^{-2} \left[ F(1/t) + \mathcal{O}(b^2) \right]} \,.
\end{equation}
This results in finite semiclassical blocks, $\mathcal{F}$. For example, for \eqref{blockat0expansion}, we get the finite quantities
\begin{equation}\label{finitesemiclassicallimit}
\FIVsc{a_\infty}{a_t}{a}{a_1}{{a_{0 \theta}}}{a_{2,1}}{a_0}{\frac{1}{t}}{z} = \lim_{b \to 0} \frac{\FIV{\alpha_\infty}{\alpha_t}{\alpha}{\alpha_1}{{\alpha_{0 \theta}}}{\alpha_{2,1}}{\alpha_0}{\frac{1}{t}}{z}}{\mathfrak{F} \left( \begin{matrix} \alpha_t \\ \alpha_\infty \end{matrix} \alpha \, \begin{matrix} \alpha_1 \\ \alpha_0 \end{matrix} ; \frac{1}{t} \right)},
\end{equation}
where recall $a_i = b \alpha_i$.
It is $\mathcal{F}$ that obeys the Heun equation, and thus the Heun connection formula for connecting $z\sim 0$ with $z\sim \infty$ is inherited from the connection formulae for $\mathfrak{F}$ discussed above, in the semiclassical limit \eqref{finitesemiclassicallimit}.

Quantities that appear in the connection coefficients derived from Liouville theory admit explicit and combinatoric expressions in terms of Young diagrams that arise in the definition of the instanton partition function of $\mathcal{N}=2$ $SU(2)$ gauge theory with fundamental matter \cite{Alday:2009aq}.
Let $Y$ denote the Young diagram with column heights $(Y_1\ge Y_2\ge\dots)$ and row lengths $(Y'_1\ge Y'_2,\dots)$. For every box $s=(i,j)$ of $Y$ we denote the arm length and the leg length as
\begin{equation}
A_Y(i, j) = Y_j -  i, \quad L_Y(i, j) =Y'_i - j.
\end{equation}
Let us denote a pair of Young diagrams with $\vec{Y}=\left( Y_1, Y_2 \right)$ and $| \vec{Y} | = | Y_1 | + | Y_2 |$ the total number of boxes. We denote the v.e.v. of the scalar in the vector multiplet with $\vec{a}=(a_1,a_2)$ and the parameters characterizing the $\Omega$-background with $\epsilon_1,\epsilon_2$. We define the hypermultiplet contribution to the partition function,
\begin{equation}
\begin{aligned}
&z_{\text{hyp}} \left( \vec{a}, \vec{Y}, m \right) =\\
&\prod_{k= 1,2} \prod_{(i,j) \in Y_k} \left[ a_k + m + \epsilon_1 \left( i - \frac{1}{2} \right) + \epsilon_2 \left( j - \frac{1}{2} \right) \right] \,, 
\end{aligned}
\end{equation}
and the vector multiplet contribution,
\begin{equation}
\begin{aligned}
&z_{\text{vec}} \left( \vec{a}, \vec{Y} \right) =\\
&\prod_{i,j=1}^2\prod_{s\in Y_i}\frac{1}{a_i-a_j-\epsilon_1L_{Y_j}(s)+\epsilon_2(A_{Y_i}(s)+1)}\\
&\prod_{t\in Y_j}\frac{1}{-a_j+a_i+\epsilon_1(L_{Y_i}(t)+1)-\epsilon_2A_{Y_j}(s)}\,.
\end{aligned}
\end{equation}
We take $\epsilon_1=1$ and $\vec{a}=(a,-a)$.
We denote with $m_1,m_2,m_3,m_4$ the masses of the four hypermultiplets and we introduce the gauge parameters $a_0,a_t,a_1,a_{\infty}$ satisfying, in our notation,
\begin{equation}\label{gaugemasses}
\begin{aligned}
m_1&=a_1-a_0,\\
m_2&=a_1+a_0,\\
m_3&=a_{\infty}+a_t,\\
m_4&=-a_{\infty}+a_t.
\end{aligned}
\end{equation}
Moreover we denote the instanton counting parameter by $1/t=e^{2\pi i\tau}$, where $\tau$ is related to the gauge coupling by $\tau=\frac{\theta}{2\pi}+i\frac{4\pi}{g_{\rm YM}^2}$. The instanton part of the Nekrasov Shatashvili free energy is then given as a power series in $1/t$ by
\begin{equation}
\begin{aligned}
F(1/t)=&\lim_{\epsilon_2\to 0}\epsilon_2\log\Biggl[\left(1-\frac{1}{t}\right)^{-2\epsilon_2^{-1}\left(\frac{1}{2}+a_1\right)\left(\frac{1}{2}+a_t\right)}\\
&\sum_{\vec{Y}}t^{-|\vec{Y}|}z_{\text{vec}} \left( \vec{a}, \vec{Y} \right)\prod_{i=1}^4z_{\text{hyp}} \left( \vec{a}, \vec{Y}, m_i \right)\Biggr].
\end{aligned}
\end{equation}
The first-order instanton expansion is given in \eqref{NSfirstinstanton}.
The parameter $a$ characterizes the composite monodromy around $z=0$ and $z=1$ in the 4-punctured sphere. From the gauge theory point of view, $a$ is expressed in a series expansion in the instanton counting parameter $1/t$, obtained by inverting the Matone relation in \eqref{Matone},
where the parameter $u$ is the complex moduli parametrizing the energy of the corresponding Seiberg Witten curve \cite{seiberg1994a,seiberg1994}, and corresponds to the accessory parameter of the Heun equation.
Explicitly, the expansion of $a$ reads as follows
\begin{equation} 
\begin{aligned}
a=&\pm\Biggl\{\sqrt{\frac{1}{4}+u-a_t^2+a_{\infty}^2}\\
&-\frac{u \left(u-a_0^2+a_1^2-a_t^2+a_{\infty}^2\right)}{2 \left(a_{\infty}^2-a_t^2+u\right) \sqrt{1+4u-4 a_t^2+4 a_{\infty}^2}}\frac{1}{t}+\mathcal{O}(1/t^2)\Biggr\}.
\end{aligned}
\end{equation}

\onecolumngrid
\subsection{Longer formulas}
The dictionary for the Heun equation reads,
\begin{equation}\label{dictiogauge}
\begin{aligned}
t&=\frac{9+\sqrt{48 \mu ^2+9}}{9-\sqrt{48 \mu ^2+9}},\\
a_0 &= \frac{\Delta}{2}-1,\\
a_1&=-\frac{i \mu  \sqrt{-\sqrt{48 \mu ^2+9}-3}}{32 \sqrt{6} \mu ^2+6 \left(3 \sqrt{32 \mu ^2+6}+\sqrt{6}\right)}\left[e \left(\sqrt{48 \mu ^2+9}+9\right)+\left(\sqrt{48 \mu ^2+9}+3\right) \hat{T} \tilde{\omega}\right],\\
a_t&=\frac{i \mu  \sqrt{\sqrt{48 \mu ^2+9}-3} }{2 \sqrt{32 \mu ^2+6} \left(\sqrt{48 \mu ^2+9}-9\right)}\left[e \left(\sqrt{48 \mu ^2+9}-9\right)+\left(\sqrt{48 \mu ^2+9}-3\right) \hat{T} \tilde{\omega}\right],\\
a_{\infty}&=\frac{i \tilde{\omega}}{4 \pi },\\
u&=\frac{\sqrt{3}}{4 \left(16 \mu ^2+3\right)^{3/2} \left(\sqrt{48 \mu ^2+9}-9\right)}\biggl[8 \mu ^4 \left(21 e^2+30 e \hat{T} \tilde{\omega}-18 \hat{k}^2+9 \hat{T}^2 \tilde{\omega}^2+32\right)\\
&+\mu ^2 \left(-9 e^2+18 e \hat{T} \tilde{\omega}-27 \hat{k}^2+9 \hat{T}^2 \tilde{\omega}^2+528\right)-3 \mu ^2 \Bigl(e^2 \sqrt{48 \mu ^2+9}+2 e \sqrt{48 \mu ^2+9} \hat{T} \tilde{\omega}\\
&+32 \sqrt{48 \mu ^2+9}-\sqrt{48 \mu ^2+9} \hat{k}^2+\sqrt{48 \mu ^2+9} \hat{T}^2 \tilde{\omega}^2\Bigr)-16 \mu ^4 \Bigl(e^2 \sqrt{48 \mu ^2+9}+2 e \sqrt{48 \mu ^2+9} \hat{T} \tilde{\omega}\\
&-\sqrt{48 \mu ^2+9} \hat{k}^2+\sqrt{48 \mu ^2+9} \hat{T}^2 \tilde{\omega}^2\Bigr)-2\Delta(\Delta-4) \left(16 \mu ^2+3\right) \left(-4 \mu ^2+\sqrt{48 \mu ^2+9}-3\right) -18 \sqrt{3} \sqrt{16 \mu ^2+3}+90\biggr].
\end{aligned}
\end{equation}
The ratio of the connection coefficients is given by
\begin{equation}\label{fullgreen}
\begin{aligned}
\frac{\mathfrak{C}_{+}^{(\infty\,0)}}{\mathfrak{C}_{-}^{(\infty\,0)}}=e^{-\partial_{a_0}F(1/t)+2i\pi a_0}\frac{\mathcal{C}^{\text{exact}}_{-a_0,a}-\mathcal{G}^{\text{exact}}\,\mathcal{C}^{\text{exact}}_{-a_0,-a}}{\mathcal{C}^{\text{exact}}_{a_0,a}-\mathcal{G}^{\text{exact}}\,
\mathcal{C}^{\text{exact}}_{a_0,-a}},
\end{aligned}
\end{equation}
with
\begin{equation}
\begin{aligned}
\mathcal{C}^{\text{exact}}_{a_0,a}&=\frac{\Gamma(2a_0)}{\Gamma\left(\frac{1}{2}+a_0+a_1+a\right)\Gamma\left(\frac{1}{2}+a_0-a_1+a\right)},\\
\mathcal{G}^{\text{exact}}&=t^{-2a}\,e^{-\partial_aF(1/t)}\,\frac{\Gamma(-2a)^2\Gamma\left(\frac{1}{2}-a_{\infty}+a_t+a\right)\Gamma\left(\frac{1}{2}-a_{\infty}-a_t+a\right)}{\Gamma(2a)^2\Gamma\left(\frac{1}{2}-a_{\infty}+a_t-a\right)\Gamma\left(\frac{1}{2}-a_{\infty}-a_t-a\right)}.
\end{aligned}
\end{equation}

The leading coefficients in the Green's function expression \eqref{Greensexpr} is given by
\be
\mathcal{C}^{(0)}_{\Delta,\nu}(\hat{k}) = \frac{\Gamma(\Delta-2)}{\Gamma\left( \frac{\Delta-1}{2} - \frac{e}{2\sqrt{3}} + \nu\right)\Gamma\left( \frac{\Delta-1}{2} + \frac{e}{2\sqrt{3}} + \nu\right)}.\label{calC0}
\ee

The subleading corrections in the Green's function expression \eqref{Greensexpr} are given by
\begin{equation}\label{C1}
\begin{aligned}
\mathcal{C}^{(1)}_{\Delta,\nu}(\tilde{\omega},\hat{k})=&\frac{\Gamma (\Delta -2)}{864 \sqrt{6} \nu  \left(1-4 \nu ^2\right) \prod_{\sigma=\pm}\Gamma \left(\frac{\Delta-1}{2}+\nu +\sigma\frac{e}{2 \sqrt{3}}\right) }\\
&\times\sum_{\sigma=\pm}\biggl[-7 \pi  e^4+7 \sqrt{6} e^3 \tilde{\omega}+3 \pi  e^2 \left(4\Delta (\Delta -4) +15 \hat{k}^2-8\right)-3 \sqrt{6} e \tilde{\omega} \left(4 \Delta(\Delta -4)  +9 \hat{k}^2-8\right)\\
&-96\,\sigma\, \nu  \left(4 \nu ^2-1\right) \left(\sqrt{3} \pi  e-3 \sqrt{2} \tilde{\omega}\right)-18 \pi  \hat{k}^2 \left(2\Delta (\Delta -4) +3 \hat{k}^2\right)\biggr] \psi ^{(0)}\left(\frac{\Delta-1 }{2}+\nu+\sigma\frac{e}{2 \sqrt{3}} \right),
\end{aligned}
\end{equation}

\begin{equation}\label{G1+}
\begin{aligned}
\mathcal{G}^{(1,+)}(\tilde{\omega},\hat{k})=&\frac{2^{\nu-2} 3^{-3 \nu-2} \pi ^{2 \nu} \Gamma (-2 \nu)^2 \Gamma \left(\nu-\frac{i e}{2 \sqrt{6}}+\frac{1}{2}\right) \Gamma \left(\frac{1}{2}-\frac{i \tilde{\omega}}{2 \pi }+\nu+\frac{i e}{2 \sqrt{6}}\right)}{2 \sqrt{6} \left(6-24 \nu^2\right) \nu \Gamma (2 \nu)^2 \Gamma \left(-\nu-\frac{i e}{2 \sqrt{6}}+\frac{1}{2}\right) \Gamma \left(\frac{1}{2}-\frac{i \tilde{\omega}}{2 \pi }-\nu+\frac{i e}{2 \sqrt{6}}\right)}\,\Bigl[7 \pi  e^4-7 \sqrt{6} e^3 \tilde{\omega}\\
&+3 \pi  e^2 \left(-4 \Delta(\Delta -4) -15 \hat{k}^2+8\right)+3 \sqrt{6}\, e\, \tilde{\omega} \left(4 \Delta(\Delta -4) +9 \hat{k}^2-8\right)+18 \pi  \hat{k}^2 \left(2 \Delta(\Delta -4)  +3 \hat{k}^2\right)\Bigr],
\end{aligned}
\end{equation}
and
\begin{equation}\label{G1-}
\begin{aligned}
&\mathcal{G}^{(1,-)}(\tilde{\omega},\hat{k})=\frac{2^{\nu-3} 3^{-3 \nu-2} \pi ^{2 \nu} \Gamma (-2 \nu)^2 \Gamma \left(\frac{1}{2}+\nu-\frac{i e}{2 \sqrt{6}}\right) \Gamma \left(\frac{1}{2}+\nu+\frac{i e}{2 \sqrt{6}}-\frac{i \tilde{\omega}}{2 \pi }\right)}{2 \sqrt{6}\,\nu\, \left(6-24 \nu^2\right)^2 \Gamma (2 \nu)^2 \Gamma \left(\frac{1}{2}-\nu-\frac{i e}{2 \sqrt{6}}\right) \Gamma \left(\frac{1}{2}-\nu+\frac{i e}{2 \sqrt{6}}-\frac{i \tilde{\omega}}{2 \pi }\right)}\times\\
&\biggl\{\pi  \left(-7 \log \left(\frac{27}{2}\right)+14 \log (\pi )-10\right) e^6+\tilde{\omega} \left(7 \log \left(\frac{27}{2}\right)-14 \log (\pi )+8\right) \sqrt{6} e^5\\
&+2 \pi  \left[3 \left(11 \log \left(\frac{27}{2}\right)-22 \log (\pi )+7\right) \hat{k}^2-12 \log \left(\frac{27}{2}\right)+24 \log (\pi )+13 (\Delta -4) \Delta  \left(\log \left(\frac{27}{2}\right)-2 \log (\pi )+2\right)+78\right] e^4\\
&-2 \sqrt{6} \tilde{\omega} \left[12 \left(\log \left(\frac{729}{4}\right)-4 \log (\pi )+1\right) \hat{k}^2-12 \log \left(\frac{27}{2}\right)+24 (\log (\pi )+3)+(\Delta -4) \Delta  \left(13 \log \left(\frac{27}{2}\right)-26 \log (\pi )+20\right)\right] e^3\\
&-3 \pi  \biggl[9 \left(7 \log \left(\frac{27}{2}\right)-14 \log (\pi )+2\right) \hat{k}^4+6 \left(-4 \log \left(\frac{27}{2}\right)+8 \log (\pi )+(\Delta -4) \Delta  \left(9 \log \left(\frac{27}{2}\right)-18 \log (\pi )+8\right)+20\right) \hat{k}^2\\
&+8 \left((\Delta -4) \Delta  \left(\log \left(\frac{4 \pi ^4}{729}\right)+(\Delta -4) \Delta  \left(\log \left(\frac{27}{2}\right)-2 \log (\pi )+3\right)+16\right)+24\right)\biggr] e^2\\
&+3 \tilde{\omega} \biggl[6 \left(-4 \log \left(\frac{27}{2}\right)+8 \log (\pi )+(\Delta -4) \Delta  \left(5 \log \left(\frac{27}{2}\right)-10 \log (\pi )+4\right)+16\right) \hat{k}^2+16 (\Delta -2)^2 ((\Delta -4) \Delta +3)\\
&+\left(-27 \hat{k}^4-8 (\Delta -4) \Delta  ((\Delta -4) \Delta -2)\right) \log \left(\frac{2 \pi ^2}{27}\right)\biggr] \sqrt{6} e+2 \pi  \left(3 \hat{k}^2+2 (\Delta -4) \Delta \right)^2 \biggl[3 \left(\log \left(\frac{19683}{8}\right)-6 \log \pi+1\right) \hat{k}^2\\
&+2 (\Delta -4) \Delta +6\biggr]+\left(6-24 \nu^2\right) \biggl[7 \pi  e^4+7 (2 i \pi  \nu-\tilde{\omega}) \sqrt{6} e^3-3 \left(\pi  \left(15 \hat{k}^2+4 (\Delta -4) \Delta -8\right)+28 i \nu \tilde{\omega}\right) e^2\\
&+\left(3 \left(9 \hat{k}^2+4 (\Delta -4) \Delta -8\right) \tilde{\omega}-14 i \pi  \left(3 \hat{k}^2+2 (\Delta -4) \Delta \right) \nu\right) \sqrt{6} e+6 \left(3 \hat{k}^2+2 (\Delta -4) \Delta \right) \left(3 \pi  \hat{k}^2+14 i \nu \tilde{\omega}\right)\biggr]\times\\
&\times\psi ^{(0)}\left(\frac{1}{2}+\frac{i e}{2 \sqrt{6}}-\nu-\frac{i \tilde{\omega}}{2 \pi }\right)+\left(6-24 \nu^2\right) \biggl[7 \pi  e^4+7 (-2 i \pi  \nu-\tilde{\omega}) \sqrt{6} e^3+3 \left(\pi  \left(-15 \hat{k}^2-4 (\Delta -4) \Delta +8\right)+28 i \nu \tilde{\omega}\right) e^2\\
&+\left(3 (14 i \pi  \nu+9 \tilde{\omega}) \hat{k}^2-24 \tilde{\omega}+4 (\Delta -4) \Delta  (7 i \pi  \nu+3 \tilde{\omega})\right) \sqrt{6} e+6 \left(3 \hat{k}^2+2 (\Delta -4) \Delta \right) \left(3 \hat{k}^2 \pi -14 i \nu \tilde{\omega}\right)\biggr]\times\\
&\times\psi ^{(0)}\left(\frac{1}{2}+\frac{i e}{2 \sqrt{6}}+\nu-\frac{i \tilde{\omega}}{2 \pi }\right)+\left(6-24 \nu^2\right) \biggl[7 \pi  e^4+7 (-2 i \pi  \nu-\tilde{\omega}) \sqrt{6} e^3+3 \left(\pi  \left(-15 \hat{k}^2-4 (\Delta -4) \Delta +8\right)+28 i \nu \tilde{\omega}\right) e^2\\
&+\left(3 (14 i \pi  \nu+9 \tilde{\omega}) \hat{k}^2-24 \tilde{\omega}+4 (\Delta -4) \Delta  (7 i \pi  \nu+3 \tilde{\omega})\right) \sqrt{6} e+6 \left(3 \hat{k}^2+2 (\Delta -4) \Delta \right) \left(3 \hat{k}^2 \pi -14 i \nu \tilde{\omega}\right)\biggr]\times\\
&\times\psi ^{(0)}\left(\frac{1}{2}-\frac{i e}{2 \sqrt{6}}-\nu\right)-4 \left(6-24 \nu^2\right) \biggl[7 \pi  e^4-7 \sqrt{6} \tilde{\omega} e^3+3 \pi  \left(-15 \hat{k}^2-4 (\Delta -4) \Delta +8\right) e^2\\
&+3 \left(9 \hat{k}^2+4 (\Delta -4) \Delta -8\right) \tilde{\omega} \sqrt{6} e+18 \hat{k}^2 \pi  \left(3 \hat{k}^2+2 (\Delta -4) \Delta \right)\biggr] \psi ^{(0)}(-2 \nu)-4 \left(6-24 \nu^2\right) \biggl[7 \pi  e^4-7 \sqrt{6} \tilde{\omega} e^3\\
&+3 \pi  \left(-15 \hat{k}^2-4 (\Delta -4) \Delta +8\right) e^2+3 \left(9 \hat{k}^2+4 (\Delta -4) \Delta -8\right) \tilde{\omega} \sqrt{6} e+18 \hat{k}^2 \pi  \left(3 \hat{k}^2+2 (\Delta -4) \Delta \right)\biggr] \psi ^{(0)}(2 \nu)\\
&+\left(6-24 \nu^2\right) \biggl[7 \pi  e^4+7 (2 i \pi  \nu-\tilde{\omega}) \sqrt{6} e^3-3 \left(\pi  \left(15 \hat{k}^2+4 (\Delta -4) \Delta -8\right)+28 i \nu \tilde{\omega}\right) e^2+\biggl(3 \left(9 \hat{k}^2+4 (\Delta -4) \Delta -8\right) \tilde{\omega}\\
&-14 i \pi  \left(3 \hat{k}^2+2 (\Delta -4) \Delta \right) \nu\biggr) \sqrt{6} e+6 \left(3 \hat{k}^2+2 (\Delta -4) \Delta \right) \left(3 \pi  \hat{k}^2+14 i \nu \tilde{\omega}\right)\biggr] \psi ^{(0)}\left(\frac{1}{2}-\frac{i e}{2 \sqrt{6}}+\nu\right)\biggr\}.
\end{aligned}
\end{equation}

The diffusivity coefficient in the hydrodynamic expansion in \eqref{hydromode} reads 
\begin{equation}\label{hydroorder}
\begin{aligned}
&D^{(n)}=-\frac{\pi \hat{T}_c^{(n)}}{8\tilde{\nu} \left[\psi ^{(0)}\left(\frac{1}{2}+\frac{i e}{2 \sqrt{6}}-\tilde{\nu}\right)-\psi ^{(0)}\left(\frac{1}{2}+\frac{i e}{2 \sqrt{6}}+\tilde{\nu}\right)\right]}\times\\
&\Biggl\{\psi ^{(0)}\left(\frac{\Delta-1}{2}+\frac{e}{2\sqrt{3}}+\tilde{\nu}\right)+\psi ^{(0)}\left(\frac{\Delta-1}{2}-\frac{e}{2\sqrt{3}}+\tilde{\nu}\right)+\psi ^{(0)}\left(\frac{\Delta-1}{2}+\frac{e}{2\sqrt{3}}-\tilde{\nu}\right)\\
&+\psi ^{(0)}\left(\frac{\Delta-1}{2}-\frac{e}{2\sqrt{3}}-\tilde{\nu}\right)+\psi ^{(0)}\left(\frac{1}{2}+\frac{i e}{2 \sqrt{6}}-\tilde{\nu}\right)+\psi ^{(0)}\left(\frac{1}{2}-\frac{i e}{2 \sqrt{6}}-\tilde{\nu}\right)+\psi ^{(0)}\left(\frac{1}{2}+\frac{i e}{2 \sqrt{6}}+\tilde{\nu}\right)\\
&+\psi ^{(0)}\left(\frac{1}{2}-\frac{i e}{2 \sqrt{6}}+\tilde{\nu}\right)-4 \psi ^{(0)}(-2 \tilde{\nu})-4 \psi ^{(0)}(2 \tilde{\nu})+2 \log \left[\frac{1}{3}\sqrt{\frac{2}{3}} \pi\, \hat{T}_c^{(n)}\right]\Biggr\},
\end{aligned}
\end{equation}

where $\tilde{\nu}\equiv \nu|_{\hat{k}=0}$. Note that $\tilde{\nu}$ and $\hat{T}_c^{(n)}$ are functions of $\Delta,e$.

\end{document}